\documentclass[twocolumn,showpacs,preprintnumbers,amsmath,amssymb]{revtex4}

\usepackage{graphicx}
\usepackage{dcolumn}
\usepackage{bm}

\newcommand{\bq}{\begin{equation}}
\newcommand{\eq}{\end{equation}}
\newcommand{\bqa}{\begin{eqnarray}}
\newcommand{\eqa}{\end{eqnarray}}
\newcommand{\nn}{\nonumber \\}

\def\be     {\begin{equation}}
\def\ee     {\end{equation}}
\def\bea        {\begin{eqnarray}}
\def\eea        {\end{eqnarray}}
\def\bnn    {\begin{eqnarray*}}
\def\enn    {\end{eqnarray*}}

\begin{document}

\title{Nature of an intermediate non-Fermi liquid state in Ge-substituted YbRh$_{2}$Si$_{2}$:
Fermionized skyrmions, Lifshitz transition, Skyrmion liquid, and
Gruneisen ratio}
\author{Ki-Seok Kim}
\affiliation{ Department of Physics, POSTECH, Hyoja-dong, Namgu,
Pohang, Gyeongbuk 790-784, Korea \\ Institute of Edge of
Theoretical Science (IES), Hogil Kim Memorial building 5th floor,
POSTECH, Hyoja-dong, Namgu, Pohang, Gyeongbuk 790-784, Korea }
\date{\today}

\begin{abstract}
We propose a skyrmion liquid state for the non-Fermi liquid (NFL)
phase in Ge-substituted YbRh$_{2}$Si$_{2}$, where skyrmions form
their Fermi surface, argued to result from the strongly coupled
nature between skyrmions and itinerant electrons. The fermionized
skyrmion theory identifies the antiferromagnetic (AF) transition
with the Lifshitz transition, where the quantum critical point
(QCP) is characterized by the dynamical critical exponent $z = 2$.
Nonlocal interactions between skyrmions allow a critical line
above the AF QCP, which originates from the Kondo-coupling effect
with itinerant electrons. This critical line is described by the
skyrmion liquid state, which results in Landau damping for spin
fluctuations, thus characterized by $z = 3$. As a result, the
Gruneisen ratio is predicted to change from $\sim T^{-1}$ at the
AF QCP to $\sim T^{-2/3}$ in the NFL phase.
\end{abstract}


\maketitle

\section{Introduction}

Recent experiments on Yb-based heavy fermion systems of
YbRh$_{2}$(Si$_{0.95}$Ge$_{0.05}$)$_{2}$ \cite{Ge_YbRh2Si2},
Yb(Rh$_{0.94}$Ir$_{0.06}$)$_{2}$Si$_{12}$ \cite{Ir_YbRh2Si2},
YbAgGe \cite{YbAgGe}, and $\beta-$YbAlB$_{4}$ \cite{YbAlB4} have
uncovered the appearance of non-Fermi liquid physics over a finite
zero$-T$ region of the magnetic field- or pressure-tuned phase
diagram, rather than at a single quantum critical point (QCP). The
implication of these experiments casts a doubt on our fundamental
understanding for correlated electrons in metals because the
present theoretical framework cannot access such a non-Fermi
liquid phase away from quantum criticality above one dimension in
the clean system \cite{Disorder_NFL_Review}. The Fermi liquid
phase has been regarded as the only fully symmetric quantum ground
state for metals away from quantum criticality above one dimension
in the absence of randomness, sometimes referred as the mother
state of metals \cite{Shankar_RG}. A non-Fermi liquid phase away
from quantum criticality, proposed to be another stable fixed
point in the renormalization group sense, is extremely difficult
to reach in two or three dimensions except for special cases such
as the fractional quantum Hall phase because accessible stable
fixed points are non-interacting like the Fermi liquid state in
our perturbative many-body theoretical framework, generally
speaking. On the other hand, the non-Fermi liquid at a QCP is
identified with an unstable (either interacting or
non-interacting) fixed point, the origin of which is the presence
of critical fluctuations associated with a quantum phase
transition \cite{HFQCP_Review}.

In this paper we revisit an antiferromagnetic quantum phase
transition in the heavy-fermion system, where an antiferromagnetic
order from localized magnetic moments becomes destroyed by the
Kondo effect from itinerant electrons. In particular, we focus on
the non-Fermi liquid phase in Ge-substituted YbRh$_{2}$Si$_{2}$,
where tuning magnetic fields allows such an intermediate state
between antiferromagnetic and heavy fermion Fermi-liquid phases
\cite{Ge_YbRh2Si2}. First, we notice the low antiferromagnetic
transition temperature at zero magnetic field. If we interpret the
antiferromagnetic transition with the condensation of skyrmion
excitations as the superconducting transition can be understood by
the condensation of vortices in the dual picture, the low Neel
temperature implies that skyrmions can be easily created. Then,
our problem can be mapped into skyrmion dynamics interacting with
itinerant electrons in the presence of the Kondo-coupling effect.
This view point reminds us of the phase diagram for type-II
superconductors in the plane of magnetic field and temperature,
where rich vortex phases emerge to cause non-Fermi liquid physics
\cite{Vortex_Phase_Diagram}. Actually, the analogy between our
skyrmion-dynamics problem and the vortex phase diagram serves an
essential physical picture in this study. A novel vortex phase was
proposed in the region above the vortex lattice state, identified
with a vortex liquid phase, where the strongly coupled dynamics
between vortices and itinerant electrons gives rise to such an
exotic liquid state \cite{Fisher_Fermionization}.

We propose a skyrmion liquid state for the nature of the non-Fermi
liquid phase in Ge-substituted YbRh$_{2}$Si$_{2}$, the
characteristic feature of which is the power-law behavior in the
skyrmion density-density correlation function. A key ingredient in
this theoretical proposal is to fermionize skyrmion excitations at
the antiferromagnetic QCP. Although this fermionization procedure
cannot be justified above one dimension in a rigorous sense a
priori, essentially the same idea has been applied to describe
non-Fermi liquid transport phenomena in the vortex liquid phase
\cite{Fisher_Fermionization}, where the Chern-Simons term, which
appears from the payment for statistical transmutation, becomes
irrelevant at criticality \cite{Fermionization_Vortex}. This
fermionic skyrmion conjecture identifies the antiferromagnetic
quantum phase transition with the Lifshitz transition for skyrmion
excitations, where the skyrmion chemical potential touches the
skyrmion band, and thus, the dynamical critical exponent is given
by $z = 2$. If we translate tuning of the magnetic field with
controlling of the skyrmion chemical potential, we are driven to
conclude that the Fermi surface of skyrmions emerges and the
Landau damping from their particle-hole excitations gives rise to
the dynamical critical exponent $z = 3$ for spin fluctuations. As
a result, we predict that the Gruneisen ratio, given by the ratio
between the thermal expansion and specific heat coefficients,
changes from $\Gamma(T) \propto T^{-1}$ at the antiferromagnetic
(Lifshitz) QCP to $\Gamma(T) \propto T^{-2/3}$ in the non-Fermi
liquid (skyrmion liquid) phase.

\section{Skyrmion liquid}

\subsection{O(3) nonlinear $\sigma$ model}

We start from an effective Kondo-Heisenberg lattice Hamiltonian,
\bqa && H = \sum_{\boldsymbol{k}} (\epsilon_{\boldsymbol{k}} -
\mu) c_{\boldsymbol{k}\sigma}^{\dagger} c_{\boldsymbol{k}\sigma} +
J_{K} \sum_{i} \boldsymbol{s}_{i} \cdot \boldsymbol{S}_{i} + J
\sum_{ij} \boldsymbol{S}_{i} \cdot \boldsymbol{S}_{j} ,
%- H \sum_{i} (\boldsymbol{S}_{i}^{z} + \boldsymbol{s}_{i}^{z})
\nn \eqa which shows the competition between antiferromagnetic
ordering ($J$) from localized spins ($\boldsymbol{S}_{i}$) and the
formation of heavy electrons from the Kondo effect ($J_K$) of
itinerant electrons ($c_{i\sigma}$) and local spins.
%$\boldsymbol{s}_{i} = \frac{1}{2} c_{i\sigma}^{\dagger} \boldsymbol{\sigma}_{\sigma\sigma'} c_{i\sigma'}$
Since we focus on the question how the antiferromagnetic order
vanishes due to the Kondo-coupling effect, the nonlinear $\sigma$
model is our good starting point for antiferromagnetism from
localized spins \cite{Si_NLsM,Jones_NLsM}. A careful point in this
mapping is that itinerant electrons couple to ferromagnetic
fluctuations of local spins dominantly because the nesting
property of the Fermi surface does not exist due to their small
Fermi surface. Their particle-hole excitations give rise to the
Landau damping dynamics for such ferromagnetic fluctuations via
the Kondo-coupling effect, given by
\begin{widetext}
\bqa && \mathcal{S}_{eff} = \mathcal{S}_{NLsM} +
\mathcal{S}_{Kondo} , ~~~~~ \mathcal{S}_{NLsM} = \frac{c_r}{2g_r}
\int_{0}^{\beta} d \tau \int d^{d} \boldsymbol{r} \Bigl\{
\frac{1}{c_r^{2}} [\partial_{\tau}
\boldsymbol{n}(\boldsymbol{r},\tau)]^{2}
%\frac{1}{c_r^{2}} [\partial_{\tau} \boldsymbol{n}(\boldsymbol{r},\tau) - H \boldsymbol{\hat{z}} \times \boldsymbol{n}(\boldsymbol{r},\tau)]^{2}
+ [\boldsymbol{\nabla} \boldsymbol{n}(\boldsymbol{r},\tau)]^{2}
\Bigr\} + \mathcal{S}_{B} , \nn && \mathcal{S}_{Kondo} = - N_{F}
\lambda^{2} \frac{1}{\beta} \sum_{i\Omega} \int \frac{d^{d}
\boldsymbol{q}}{(2\pi)^{d}}
\Bigl(\boldsymbol{n}(\boldsymbol{r},\tau) \times [\partial_{\tau}
\boldsymbol{n}(\boldsymbol{r},\tau)]
\Bigr)_{\boldsymbol{q},i\Omega} \frac{|\Omega|}{v_{F}
|\boldsymbol{q}|} \Bigl(\boldsymbol{n}(\boldsymbol{r}',\tau')
\times [\partial_{\tau'} \boldsymbol{n}(\boldsymbol{r}',\tau')]
\Bigr)_{-\boldsymbol{q},-i\Omega} . \eqa
\end{widetext} $\mathcal{S}_{NLsM}$ is the typical nonlinear
$\sigma$ model for the Heisenberg-type model \cite{NLsM}, where
the coupling constant $g_{r}$ between antiferromagnetic spin waves
and the spin-wave velocity $c_{r}$ are renormalized by the
Kondo-coupling term. $\mathcal{S}_{B}$ denotes a single-spin Berry
phase term, which originates from the path-integral quantization
of a spin field in the spin-coherent state representation. In the
present study we will not take into account the role of this Berry
phase term. $\mathcal{S}_{Kondo}$ describes an effective
interaction between ferromagnetic fluctuations with $\lambda =
J_{K} / J$, where the polarization kernel
$\Pi(\boldsymbol{q},i\Omega) = \frac{|\Omega|}{v_{F}
|\boldsymbol{q}|}$ shows the Landau damping dynamics.

An important remark in this setup is that the Landau damping form
seems to be robust beyond the present one-loop approximation
\cite{Beyond_Eliashberg_Lee,Beyond_Eliashberg_Max,Beyond_Eliashberg_KS},
where the existence itself of the Fermi surface seems to protect
the Landau damping form. However, we would like to point out that
the exact expression for the polarization kernel still remains as
an open question. Within this uncertainty, our problem is
clarified as follows. What is the role of the nonlocal effective
interaction for ferromagnetic spin fluctuations
($\mathcal{S}_{Kondo}$) in the antiferromagnetic quantum phase
transition described by the nonlinear $\sigma$ model
($\mathcal{S}_{NLsM}$)?

%In the present study we try to answer this fundamental question,
%considering a general expression for the polarization kernel.

\subsection{Duality transformation}

Performing the duality transformation for this effective nonlinear
$\sigma$ model, we derive an effective field theory for skyrmion
excitations \cite{DQCP}. First, we rewrite Eq. (2) in terms of
bosonic spinons $z_{i\sigma}$, given by
$\boldsymbol{n}(\boldsymbol{r},\tau) = \frac{1}{2}
z_{\sigma}^{\dagger}(\boldsymbol{r},\tau)
\boldsymbol{\sigma}_{\sigma\sigma'}
z_{\sigma'}(\boldsymbol{r},\tau)$. Second, we consider an
easy-plane approximation, given by $z_{\sigma} =
\frac{1}{\sqrt{2}} e^{i\phi_{\sigma}}$, where the O(3) symmetry is
reduced to O(2).
%
%Although this symmetry reduction can give rise to unexpected
%errors sometimes, for example, resulting in weak first order
%transitions instead of expected second order ones \cite{DQCP_N},
%such a reduced symmetric model has been manipulated to allow
%qualitatively similar features, compared with the original
%symmetric model. In particular, deconfined quantum criticality has
%been predicted within this symmetry-reduced model, which also
%appears in the full symmetric model \cite{DQCP}.
%
Resorting to this CP$^{1}$ representation with the easy plane
anisotropy, we obtain
\begin{widetext}
\bqa &&
%Z = \int D \phi_{\sigma}(\boldsymbol{r},\tau) D a_{\mu}(\boldsymbol{r},\tau) e^{- \int_{0}^{\beta_r} d \tau \int d^{2} \boldsymbol{r} \mathcal{L}_{eff}}, \nn &&
\mathcal{S}_{eff} = \int_{0}^{\beta_r} d \tau \int d^{2}
\boldsymbol{r} \Bigl\{ \frac{1}{2g_r} (\partial_{\mu}
\phi_{\sigma} - a_{\mu})^{2} + \frac{1}{2e_{a}^{2}} (\partial
\times \boldsymbol{a})^{2} \Bigr\}
%- \frac{H}{g_{r}c_{r}} [\partial_{\tau} \phi_{\uparrow}(\boldsymbol{r},\tau) - \partial_{\tau} \phi_{\downarrow}(\boldsymbol{r},\tau)] \nn &&
\nn && - \int_{0}^{\beta_r} d \tau \int d^{2} \boldsymbol{r}
\int_{0}^{\beta_r} d \tau' \int d^{2} \boldsymbol{r}' c_{r} N_{F}
\lambda^{2} [\partial_{\tau} \phi_{\uparrow}(\boldsymbol{r},\tau)
- \partial_{\tau} \phi_{\downarrow}(\boldsymbol{r},\tau)]
\Pi(\boldsymbol{r}-\boldsymbol{r}',\tau-\tau') [\partial_{\tau'}
\phi_{\uparrow}(\boldsymbol{r}',\tau') -
\partial_{\tau'} \phi_{\downarrow}(\boldsymbol{r}',\tau')] ,
\eqa
\end{widetext}
where $e_{a}$ is an internal electric charge for bosonic spinons,
which results from the U(1) gauge redundancy in the CP$^{1}$
representation, and $\tau$ is scaled into $c_{r} \tau$ with
$\beta_{r} = c_{r} \beta$.

It is straightforward to perform the duality transformation for
Eq. (3) although the Kondo-fluctuation induced term gives rise to
complications. The resulting skyrmion field theory is given by
\begin{widetext}
\bqa &&
%Z_{sk} = \int D \Phi_{s} D c_{\mu} e^{-\int_{0}^{\beta} d \tau \int d^{2} r {\cal L}_{eff}} , \nn &&
{\cal S}_{sk} = \int_{0}^{\beta_r} d \tau \int d^{2}
\boldsymbol{r} \Bigl\{ \mu_{sk} \Phi_{s}^{\dagger}
(\partial_{\tau} - i c_{\tau}) \Phi_{s} + |(\partial_{\mu} - i
c_{\mu} ) \Phi_{s}|^{2} + m_{s}^{2}|\Phi_{s}|^{2} +
\frac{u_{s}}{2} |\Phi_{s}|^{4} + \frac{1}{2q_{r}^{2}} (\partial
\times \boldsymbol{c})^{2} \Bigr\} \nn && + \int_{0}^{\beta_r} d
\tau \int d^{2} \boldsymbol{r} \int_{0}^{\beta_r} d \tau' \int
d^{2} \boldsymbol{r}' \frac{1}{2 p_{r}^{2}} [\partial \times
\boldsymbol{c}(\boldsymbol{r},\tau)]_{\tau}
\Pi_{\boldsymbol{r}\boldsymbol{r}',\tau\tau'} [\partial \times
\boldsymbol{c}(\boldsymbol{r}',\tau')]_{\tau} , \eqa
\end{widetext}
where $\Phi_{s}(\boldsymbol{r},\tau)$ represents a skyrmion field
and $c_{\mu}(\boldsymbol{r},\tau)$ expresses a spin-wave
excitation. $\frac{g_r}{2} \longrightarrow \frac{1}{q_{r}^{2}}$
and $\frac{2 c_{r} N_{F} g_{r}^{2} \lambda^{2}}{v_{F}}
\longrightarrow \frac{1}{p_{r}^{2}}$ have been done to translate
two coupling constants, $J$ and $J_{K}$ of the original model into
two kinds of internal charges of skyrmions, $q_{r}$ and $p_{r}$.
$\Pi_{\boldsymbol{r}\boldsymbol{r}',\tau\tau'}$ represents the
effect from Fermi surface fluctuations, given by
$\Pi(\boldsymbol{q},i\Omega) = |\Omega|/|\boldsymbol{q}|$ in the
momentum-frequency space. We would like to emphasize that our
formulation can be constructed in any general expression for the
polarization kernel. In this respect we can give an answer beyond
some limited approximations for the polarization kernel. See
appendix A for the derivation from Eq. (3) to Eq. (4).

Several remarks should be given for our effective field theory
[Eq. (4)] of skyrmions. Since two species of bosonic spinons exist
in the CP$^{1}$ representation, there must be two kinds of
vortices, $\Phi_{\uparrow}$ and $\Phi_{\downarrow}$, identified
with meron excitations \cite{DQCP}. However, we take the limit of
$e_{a} \rightarrow \infty$, resulting in the fact that all gauge
non-singlet excitations are confined to disappear from the
effective field theory \cite{DHLee,Nayak}. As a result, $\Phi_{s}
\sim \Phi_{\uparrow} \Phi_{\downarrow}^{\dagger}$ arises naturally
from the confinement ansatz of $e_{a} \rightarrow \infty$,
identified with the skyrmion field instead of meron excitations.
%There is no room for exotica such as deconfined quantum criticality in this setup.
The effective field theory, Eq. (4) consists of ``normal"
skyrmions and spin wave excitations in the easy plane
approximation, where the U(1) gauge symmetry is associated with
the conservation of the $z-$component of localized spins. Another
important aspect is the presence of the particle-hole
symmetry-breaking term, given by the linear time-derivative with
$\mu_{sk}$ in the skyrmion dynamics. As discussed in the
introduction, our physical picture is based on the analogy with
the vortex phase diagram of type II superconductors
\cite{Vortex_Phase_Diagram}. We speculate that there exists a
skyrmion ``lattice"$-$like region at least at finite temperatures
before the antiferromagnetic order disappears, which breaks the
particle-hole symmetry for skyrmion dynamics as the vortex lattice
phase. Appearance of the particle-hole symmetry breaking is
attributed to the low Neel temperature and the Kondo-coupling
effect with itinerant electrons \cite{JH_Skyrmion}. Our physical
picture is that skyrmions serve effective magnetic fields to
itinerant electrons, which can play a role in reducing the kinetic
energy of itinerant electrons. As a result, the total energy can
be more lowered, where skyrmion fluctuations become more softened
by the interplay between itinerant electrons and skyrmions.
Indeed, a similar phenomenon has been observed theoretically in
the Kondo system on geometrically frustrated lattices, where
coplanar ordering of localized spins in the absence of itinerant
electrons becomes unstable to turn into spin chiral ordering
structures at special fillings of itinerant electrons, which
generate effective magnetic fields to itinerant electrons and
quench the kinetic energy to lower the total ground-state energy
\cite{Kondo_Frustration}. The antiferromagnetic QCP is achieved by
proliferation of skyrmion excitations, but we claim that there can
exist an intermediate skyrmion liquid state before the skyrmion
condensed phase.

%\bqa && \Phi_{bs} = \sqrt{Z_{\Phi}} \Phi_{rs} , ~~~
%\boldsymbol{c}_{b} = \sqrt{Z_{c}} \boldsymbol{c}_{r} , ~~~
%m_{bs}^{2} = Z_{\Phi}^{-1} Z_{m} m_{rs}^{2} , \nn && u_{bs} =
%\mu^{\epsilon} Z_{\Phi}^{-2} Z_{u} u_{rs} , ~~~ q_{b}^{-2} =
%\mu^{-\epsilon} Z_{c} q_{r}^{-2} , ~~~ p_{b}^{-2} =
%\mu^{-\epsilon} Z_{c} p_{r}^{-2} \nonumber \eqa

An interesting feature of Eq. (4) is the nonlocal term with the
polarization kernel in the gauge or spin-wave propagator. This
gives rise to the fact that the ``charge" $p_{r}$ does not
renormalize at the QCP when the linear time-derivative term is
neglected.
%
%The gauge invariance results in the renormalization group
%relations between bare and renormalized coupling constants,
%$q_{b}^{-2} = \mu^{-\epsilon} Z_{c} q_{r}^{-2}$ and $p_{b}^{-2} = \mu^{-\epsilon} Z_{c} p_{r}^{-2}$
%
It is straightforward to obtain renormalization group equations
for two coupling constants, \bqa && \frac{d q_{r}^{2}}{d \ln \mu}
= q_{r}^{2} (\epsilon - \eta_{c}) , ~~~~~ \frac{d p_{r}^{2}}{d \ln
\mu} = p_{r}^{2} (\epsilon - \eta_{c}) . \eqa $\eta_{c} = \frac{d
\ln Z_{c}}{d \ln \mu} = \mathcal{C}_{c} q_{r}^{2}$ is the
anomalous dimension of the gauge field, where $Z_{c}$ is the
wave-function renormalization constant for the gauge field and
$\mathcal{C}_{c}$ is a positive numerical constant. $\epsilon
\equiv 3 - d$ is with dimension $d$. A key point in these
renormalization group equations is that the anomalous dimension
$\eta_{c}$ appears in the same way, which originates from the
gauge invariance. As a result, we obtain $\eta_{c} = \epsilon$ at
the charged fixed point, which is exact for all orders
\cite{Tesanovic_Herbut}. This defines the critical coupling
constant $q_{r}^{c}$. However, there is no equation to fix
$p_{r}$, which remains unrenormalized. This originates from the
nonlocal interaction generated from the Kondo-coupling effect. The
non-renormalization for $p_{r}$ motivates us to search an
appropriate description for the existence of a critical line
instead of a critical point. One way is to fermionize skyrmion
excitations. The fermionization procedure gives rise to
non-relativistic dynamics for such fermionized skyrmion
excitations, characterized by $z = 2$. The $z = 2$ dynamics seems
to be inconsistent with the bosonic description with $z = 1$. Our
requirement of mathematical consistency between fermionic and
bosonic descriptions for skyrmions forces us to introduce the
linear time-derivative term in Eq. (4). The physical picture for
this particle-hole symmetry breaking term was discussed in the
previous paragraph.

%$D_{\tau\tau}(\boldsymbol{q},i\Omega) = \frac{q_{r}^{2}}{q^{2}}$
%$D_{ij}(\boldsymbol{q},i\Omega) = \frac{1}{\frac{\Omega^{2} + q^{2}}{q_{r}^{2}} + \frac{|\Omega| q}{ p_{r}^{2}}} \Bigl( \delta_{ij} - \frac{q_{i}q_{j}}{q^{2}} \Bigr)$

%$\frac{d \ln Z_{\Phi}}{d \ln \mu} = - p_{r}^{2} f(p_{r}^{2}/q_{r}^{2})$
%$f(p_{r}^{2}/q_{r}^{2}) = \frac{1}{(2\pi)^{d}} \int_{0}^{\infty} d x x^{d-1} \frac{1}{1 + x^{2}} \frac{1}{(p_{r}^{2}/q_{r}^{2}) (1 + x^{2}) + x}$
%$\frac{d \ln Z_{m}}{d \ln \mu} = - \mathcal{C}_{m} u_{rs}$

%$\frac{1}{\nu(p_{r})} = 2 - \mathcal{C}_{m} u_{rs}^{c}(p_{r}) - p_{r}^{2} f[(p_{r}/q_{r}^{c})^{2}]$ $\frac{1}{\nu(p_{r} \rightarrow \infty)} \leq \frac{1}{\nu(p_{r})} \leq \frac{1}{\nu(p_{r} \rightarrow 0)}$

%\bqa && \frac{d m_{rs}^{2}}{d \ln \mu} = \Bigl( 2 - \mathcal{C}_{m} u_{rs} - p_{r}^{2} f(p_{r}^{2}/q_{r}^{2}) \Bigr) m_{rs}^{2} \eqa

%$\frac{d \ln Z_{u}}{d \ln \mu} = - \mathcal{C}_{u} u_{rs} + p_{r}^{2} g(p_{r}^{2}/q_{r}^{2})$

%\bqa && \frac{d u_{rs}}{d \ln \mu} = \Bigl( \epsilon + p_{r}^{2} [2f(p_{r}^{2}/q_{r}^{2})+g(p_{r}^{2}/q_{r}^{2})] \Bigr) u_{rs}
%- \mathcal{C}_{u} u_{rs}^{2} - p_{r}^{2} h(p_{r}^{2}/q_{r}^{2}) \eqa

%$g(p_{r}^{2}/q_{r}^{2}) = \frac{4}{(2\pi)^{2d+1}} \int_{0}^{\infty} d z z \int_{0}^{\infty} d y \int_{0}^{2\pi} d \theta
%\int_{0}^{\infty} d x x^{d-1} \frac{(q_{r}^{2}/p_{r}^{2}) z^{2} (- y^{2} + 1/4) + \frac{x^{2} \sin^{2}
%\theta}{(p_{r}^{2}/q_{r}^{2}) (z^{2} + 1) + z }}{[z^{2} (y^{2} + 1) + (x^{2} + 1 + 2 x \cos \theta)][z^{2} y^{2} + x^{2}]}$

\subsection{Fermionized skyrmions and Lifshitz transition}

Possibility of statistical transmutation for vortices has been
discussed in the vortex liquid phase \cite{Fisher_Fermionization}
and at quantum criticality in geometrically frustrated spin
systems \cite{Fermionization_Vortex}. Fermionization of skyrmion
excitations can be performed in the same way as that of vortices.
The key point is that the Chern-Simons term becomes irrelevant at
quantum criticality
\cite{Fisher_Fermionization,Fermionization_Vortex}, implying that
the quantum statistics for skyrmions or vortices may not be well
defined at quantum criticality. Actually, such excitations are
strongly interacting at criticality, where we are not allowed to
pin down elementary excitations clearly. See appendix B for
irrelevance of the Chern-Simons term at quantum criticality. We
reach an effective field theory of fermionic skyrmions,
\begin{widetext}
\bqa &&
%Z_{sk} = \int D \Phi_{s} D c_{\mu} e^{-\int_{0}^{\beta} d \tau \int d^{2} r {\cal L}_{eff}} , \nn &&
{\cal S}_{eff} = \int_{0}^{\beta_r} d \tau \int d^{2}
\boldsymbol{r} \Bigl\{ \psi_{s}^{\dagger} (\partial_{\tau} -
\mu_{\psi} - i c_{\tau}) \psi_{s} + \frac{1}{2m_{sk}}
|(\boldsymbol{\nabla} - i \boldsymbol{c} ) \psi_{s}|^{2} +
\frac{1}{2q_{r}^{2}} (\partial \times \boldsymbol{c})^{2} \Bigr\}
\nn && + \int_{0}^{\beta_r} d \tau \int d^{2} \boldsymbol{r}
\int_{0}^{\beta_r} d \tau' \int d^{2} \boldsymbol{r}' \frac{1}{2
p_{r}^{2}} [\partial \times
\boldsymbol{c}(\boldsymbol{r},\tau)]_{\tau}
\Pi_{\boldsymbol{r}\boldsymbol{r}',\tau\tau'} [\partial \times
\boldsymbol{c}(\boldsymbol{r}',\tau')]_{\tau} , \eqa
\end{widetext}
where $\psi_{s}$ represents the fermionic skyrmion field with the
skyrmion chemical potential $\mu_{\psi}$ and the band mass $m_{sk}
\propto \mu_{sk}$. The antiferromagnetic transition is described
by tuning the chemical potential from $\mu_{\psi} < 0$ to
$\mu_{\psi} > 0$. In this respect the antiferromagnetic QCP is
realized at $\mu_{\psi} = 0$, which means that the chemical
potential touches the lowest position of the band, thus identified
with the Lifshitz transition. The Lifshitz transition is
characterized by the dynamical critical exponent $z = 2$, where
the renormalization effect from gauge fluctuations is not
relevant. As a result, both $z = 2$ critical skyrmion fluctuations
and $z = 1$ gapless spin fluctuations coexist to contribute to
thermodynamics. Increasing the skyrmion chemical potential from
zero to positive, the skyrmion Fermi surface will arise. Then,
Landau damping for gauge fluctuations emerges from particle-hole
excitations of fermionic skyrmions, resulting in the $z = 3$
dynamics \cite{Lee_Nagaosa}. The renormalization group analysis
has been performed \cite{U1GT_RG,U1GT_RG_EFT}, where the
renormalization group equation for $q_{r}$ remains essentially the
same as that of Eq. (5). An effective field theory has been
constructed right at the interacting charged fixed point in the
absence of $p_{r}$ \cite{U1GT_RG_EFT}, where the coupling constant
$p_{r}$ can be easily shown to be irrelevant due to the $z = 3$
gauge dynamics. {\it Irrelevance of $p_{r}$ is consistent with
enhancement of the Kondo-coupling effect in terms of original
variables, implying that the skyrmion liquid with a skyrmion Fermi
surface is a possible solution which originates from the interplay
between itinerant electrons and skyrmion fluctuations.}

%An essential structure of the renormalization group equations of
%Eq. (5) still holds in the presence of the skyrmion Fermi surface
%\cite{U1GT_RG,U1GT_RG_EFT}, which results in non-renormalization
%of $p_{r}$, identifying the non-Fermi liquid phase with a critical
%line of a skyrmion liquid state.

\subsection{Experimental implication: Gruneisen ratio}

Although the skyrmion liquid state can be characterized by the
power-law correlation function of skyrmion densities, there are
more directly measurable thermodynamic quantities. Both specific
heat and thermal expansion coefficients can be deduced from two
critical exponents, the dynamical critical exponent $z$ and the
correlation length exponent $\nu$. Unfortunately, these two
thermodynamic functions depend on dimensionality. Although our
skyrmion description was constructed in two dimensions, we cannot
exclude the possible existence of the skyrmion liquid state in
three dimensions. In this respect it is more reliable to consider
a dimension-independent quantity, which is the Gruneisen ratio
$\Gamma(T)$, given by the ratio between the thermal expansion and
specific heat coefficients.
%
%$\gamma(T) \propto - \ln T$
%
It is straightforward to see $\Gamma(T) \propto T^{-1/\nu z}$ from
the scaling argument \cite{GR_Scaling}.

\begin{table}[ht]
\begin{tabular}{cccc}
\;\; & $z$  \; \;  & $\nu$ \; \; & $\Gamma(T) \propto T^{-1/\nu
z}$ \; \; \nn \hline AF-QCP \;\; & $2$ \; \; \; & $1/2$ \; \; \; &
$T^{-1}$ \; \; \; \nn NFL \;\; & $3$ \; \; \; & $1/2$ \; \; \; &
$T^{-2/3}$ \; \; \; \nn\hline
\end{tabular}
\caption{Gruneisen ratio at the antiferromagnetic quantum critical
point (AF-QCP) and in the non-Fermi liquid phase (NFL)}
\end{table}

Table I shows how the Gruneisen ratio changes from the
antiferromagnetic QCP to the skyrmion liquid phase. As the
dynamical critical exponent changes from $z = 2$ at the
antiferromagnetic QCP (Lifshitz transition) to $z = 3$ in the
non-Fermi liquid phase (skyrmion liquid with its Fermi surface),
the Gruneisen ratio will follow from $\sim T^{-1}$ to $\sim
T^{-2/3}$, respectively. We recall that the $T^{-1}$ behavior was
seen from CeNi$_{2}$Ge$_{2}$, regarded as a conventional
antiferromagnet, while the $T^{-2/3}$ behavior was observed from
YbRh$_{2}$(Si$_{0.95}$Ge$_{0.05}$)$_{2}$, expected to belong to a
different class of heavy fermion systems \cite{GR_Exp}. In
particular, the $T^{-2/3}$ behavior has been attributed to the
Kondo breakdown mechanism, which differs from the present scenario
\cite{GR_Kim,Kim_Jia_SF}.

\section{Discussion}

\subsection{Role of Berry phase}

The role of Berry phase has been discussed intensively in
insulating antiferromagnets \cite{DQCP}. Although it does not play
any important role in an antiferromagnetic phase, it assigns a
nontrivial quantum number to skyrmion excitations in a
paramagnetic phase, the condensation of which results in
translational symmetry breaking (valence bond solids).
Furthermore, interplay between Berry phase and interaction has
been proposed to allow exotic excitations at a quantum critical
point between an antiferromagnetic phase and a translational
symmetry broken paramagnetic state. One dimensional spin-liquid
features have been claimed to appear in two dimensions, where
spin-$1$ excitations in both antiferro- and para- magnetic phases
become fractionalized into spin-$1/2$ spinon excitations. However,
the role of Berry phase has not been clarified in metallic
antiferromagnets. We believe that this problem deserves to
investigate intensively near future.

\subsection{On the easy-plane approximation}

To take the easy-plane limit reduces the O(3) spin symmetry to
O(2) $\times$ Z$_{2}$. Although it is certainly meaningful to
study this type of antiferromagnets coupled to itinerant
electrons, an actual question is how much different physics will
arise.

First of all, we would like to point out that qualitatively
distinguished physics between O(3) and O(2) symmetries are not
kept in the present scheme of approximation, where our
calculational tool is basically based on the scheme of the $1/N$
expansion. Since critical exponents depend on $N$, one may claim
that the quantum critical point of $N = 2$ differs from that of $N
= 3$. However, this difference is just quantitative in the $1/N$
approximation. Both cases of $N = 2$ and $N = 3$ show continuous
phase transitions. On the other hand, if one claims the emergence
of fractionalized excitations in the O(N) vector model, possible
effective field theories will be given in terms of fractionalized
excitations (spinons) and gauge fluctuations \cite{DQCP}. In this
case the symmetry reduction can give rise to unexpected errors
sometimes, for example, resulting in weak first order transitions
instead of expected second order ones \cite{DQCP_N}. However, such
a reduced symmetric model can be manipulated to allow
qualitatively similar features, compared with the original
symmetric model. In particular, deconfined quantum criticality has
been predicted within the symmetry-reduced model, which also
appears in the full symmetric model \cite{DQCP}. We would like to
emphasize that these exotic excitations are not taken into account
in the present study. Instead, we consider ''conventional" spin
$1$ excitations only. In this respect we expect no qualitative
different physics between fully symmetric and symmetry reduced
spin models.

\subsection{More on particle-hole symmetry breaking in skyrmion dynamics}

It is our essential proposal to replace the second-time derivative
term with the linear-time derivative. Unfortunately, we cannot
prove the emergence of the Galileian invariance (the linear-time
derivative) from the relativistic invariance (the second-time
derivative) at present. In this respect one can call the emergence
of the Galileian invariance in the presence of itinerant electrons
as our speculation.

There must be an underlying physical mechanism for this
linear-time derivative term, which breaks the particle-hole
symmetry for ''bosonic" skyrmion excitations. It has been
demonstrated that the density of skyrmion excitations is finite at
an antiferromagnetic quantum critical point without itinerant
electrons, i.e. in an insulating antiferromagnet
\cite{Skyrmion_RG}. Of course, only the second-time derivative
term is allowed in this case, i.e., particle-hole symmetric, which
means that an equal number of skyrmion and anti-skyrmion
excitations exists at this quantum critical point. Our problem is
what happens on the particle-hole symmetry in skyrmion excitations
at the quantum critical point where itinerant charge carriers are
introduced. This is a long-standing problem, where
non-perturbative effects from interactions between itinerant
electrons and ''many" topologically nontrivial excitations should
be taken into account on equal footing. Frankly speaking, we do
not have any reliable mathematical tools for the description of
such interactions.

Our speculation is that the presence of itinerant electrons will
induce the particle-hole symmetry breaking in the skyrmion sector
because itinerant electrons favor skyrmion excitations (or
anti-skyrmions, i.e., one of the two). The physical mechanism is
as follows. When itinerant electrons move in the background of
skyrmions, they feel an effective magnetic flux, which quenches
the kinetic energy of electrons. Our expectation is that the gain
in the electron kinetic energy contribution is larger than the
cost in the particle-hole symmetry breaking of the skyrmion
sector. As a result, the particle-hole symmetry breaking is
favorable in the respect of the total energy. Actually, this
mechanism has been realized in the system of frustrated magnets,
where the presence of itinerant electrons leads the co-planar
ordering in the triangular antiferromagnet to be ordered into an
out-of-plane way, which corresponds to a spin chiral order
\cite{Kondo_Frustration}. In the present study we do not consider
this time reversal symmetry breaking because our skyrmion
excitations are strongly fluctuating at our quantum critical
point, which will not form such an exotic static order.

\subsection{Summary of approximations}

It is necessary to summarize our approximations for the procedure
from Eq. (1) to Eq. (6). It is basically an exact procedure to
derive an effective O(3) nonlinear $\sigma$ model [Eq. (2)] from
the Kondo-Heisenberg lattice Hamiltonian [Eq. (1)] after
integrating over itinerant-electron degrees of freedom if we
replace the Landau damping form for the polarization kernel with
its general expression. Although we keep the Berry phase term in
Eq. (2), we do not take into account its role in spin dynamics,
particulary, skyrmion dynamics. The easy-plane approximation has
been resorted for the derivation from Eq. (2) to Eq. (3). The
effective dual Lagrangian [Eq. (4)] in terms of bosonic skyrmions
has been found from the duality transformation, as shown in
appendix A. An essential aspect is to replace the second-time
derivative term with the linear-time derivative. This is our key
proposal. The resulting field theory [Eq. (6)] in terms of
fermionic skyrmion excitations come from the bosonic skyrmion
theory [Eq. (4)], performing the fermionization procedure. In
appendix B, we prove irrelevance of the Chern-Simons term at
quantum criticality, which supports the emergence of fermionic
skyrmions.

\subsection{On electron dynamics}

It is not straightforward to understand the feedback effect from
skyrmion excitations to itinerant electrons because we need to
calculate spin-spin correlation functions of ferromagnetic
fluctuations and this is not easy to perform, in particular, for
transverse spin fluctuations. The spin-correlation function for
the $z-$component is basically given by the flux correlator in
terms of $(\boldsymbol{\nabla} \times \boldsymbol{c})_{\tau}$.
However, the spin-correlation function for transverse components
is related with ``magnetic monopole" correlations
\cite{Fermionization_Vortex} which changes $2\pi$-flux in
$(\boldsymbol{\nabla} \times \boldsymbol{c})_{\tau}$ because
transverse components flip the $z-$component spin and the
$z-$component spin is identified with the gauge flux in the dual
representation.

%Several recent studies also focused on the question how the
%antiferromagnetic order disappears from the Kondo-coupling effect.
%These are based on the spin-wave analysis, which did not take into
%account skyrmion fluctuations. In ref. \cite{Si_NLsM} the authors
%showed the stability of the Fermi liquid phase with a small Fermi
%surface deep inside the antiferromagnetic order, where the Kondo
%coupling becomes irrelevant. Ref. \cite{Jones_NLsM} is rather in
%parallel with our study, where the authors investigated an
%effective Hertz-Moriya-Millis-type theory, integrating over
%itinerant electrons. The Kondo-coupling effect gives rise to
%nonlocal interactions between spin-wave excitations. The
%renormalization group analysis classified various fixed points.
%The QCP in their analysis was shown to be characterized by the
%dynamical critical exponent $z = 1$, consistent with our skyrmion
%description, where the spin-wave sector is given by $z = 1$ while
%the skyrmion part is described by $z = 2$.

\subsection{Comparison with a non-Fermi liquid metallic phase in the slave-fermion theory}

We would like to point out that a possible non-Fermi liquid
metallic state has been proposed in the context of the
slave-fermion theory \cite{Kim_Jia_SF,YB_ACL}. However, the
non-Fermi liquid state from the slave-fermion theory is
characterized by the existence of spin gap while the skyrmion
liquid state does not exhibit such a spin gap, where spin
fluctuations remain gapless. The skyrmion liquid phase should be
distinguished from the spin-gapped non-Fermi liquid state.

%We speculate that the heavy-fermion Fermi liquid phase may be
%described by the condensation of magnetic monopoles, which also
%force skyrmions to condense, because the proliferation of magnetic
%monopoles corresponds to infinite spin-flip processes, expected to
%catch the Kondo effect. Although the formation of the
%heavy-fermion band is not clearly understood within this dual
%description yet, it is an interesting open problem how to connect
%the scenario of monopole condensation with the Kondo breakdown theory.

\section{Conclusion}

In conclusion, we proposed a quantum critical metallic state for a
non-Fermi liquid phase of Ge-substituted YbRh$_{2}$Si$_{2}$,
identified with a skyrmion liquid state with a skyrmion Fermi
surface. Nonlocal interactions, originated from Kondo
fluctuations,
%between itinerant electrons and ferromagnetic spin excitations,
and fermionized skyrmions, argued to be allowed from both nonlocal
interactions and coexistence with itinerant electrons
\cite{Fisher_Fermionization,Fermionization_Vortex}, give rise to a
stable interacting fixed line. The antiferromagnetic transition
has been interpreted by the Lifshitz transition of fermionized
skyrmions, characterized by the dynamical exponent $z = 2$. The
skyrmion liquid state is described by its Fermi surface, resulting
in $z = 3$ for collective spin fluctuations. We suggested the
fingerprint of this scenario from the change of the Gruneisen
ratio, where it behaves from $\sim T^{-1}$ at the
antiferromagnetic quantum critical point ($z = 2$) to $\sim
T^{-2/3}$ in the non-Fermi liquid phase ($z = 3$).

An important aspect of the present study is on the introduction of
Fermi-surface fluctuations. Generally speaking, it is not
justified to integrate over Fermi-surface fluctuations. An
interesting point is that although we cannot estimate the precise
form of the polarization kernel, which results from Fermi-surface
fluctuations to give nonlocal interactions between collective spin
fluctuations, such nonlocal interactions can be irrelevant if we
assume that skyrmion excitations become fermions instead of bosons
with particle-hole symmetry breaking.

The next question will be, "Can we derive this physics based on a
purely diagrammatic way?" The present result can be translated
into the fact that nonlocal interactions from Fermi-surface
fluctuations may cause interesting novel physics to dynamics of
spin fluctuations. Although this issue has been addressed in the
duality picture, it is an important direction to prove the
emergence of $z = 3$ antiferromagnetic quantum criticality based
on the original description for spins.

\section*{Acknowledgement}

KS was supported by the National Research Foundation of Korea
(NRF) grant funded by the Korea government (MEST) (No.
2012000550).

\begin{widetext}

\appendix

\section{Duality transformation}

Starting from Eq. (3) and performing the Hubbard-Stratonovich
transformation, we obtain \bqa && Z = \int D
\phi_{\sigma}(\boldsymbol{r},\tau) D a_{\mu}(\boldsymbol{r},\tau)
D J_{\mu}^{\sigma}(\boldsymbol{r},\tau) D
j_{\tau}(\boldsymbol{r},\tau) \exp \Bigl[ - \int_{0}^{\beta_r} d
\tau \int d^{2} \boldsymbol{r} \Bigl\{ \frac{g_r}{2}
J_{\mu}^{\sigma 2} - i J_{\mu}^{\sigma} (\partial_{\mu}
\phi_{\sigma} - a_{\mu})+ \frac{1}{2e_{a}^{2}} (\partial \times
\boldsymbol{a})^{2} \nn && - i j_{\tau} (\partial_{\tau}
\phi_{\uparrow} - \partial_{\tau} \phi_{\downarrow}) \Bigr\} +
\int_{0}^{\beta_r} d \tau \int d^{2} \boldsymbol{r}
\int_{0}^{\beta_r} d \tau' \int d^{2} \boldsymbol{r}'
\frac{1}{4c_{r} N_{F} \lambda^{2}} j_{\tau}(\boldsymbol{r},\tau)
[\Pi(\boldsymbol{r}-\boldsymbol{r}',\tau-\tau')]^{-1}
j_{\tau}(\boldsymbol{r}',\tau') \Bigr] , \eqa where
$J_{\mu}^{\sigma}$ and $j_{\tau}$ are auxiliary fields which
correspond to currents.

Separating the phase field into spin-wave (smooth) and vortex
(singular) parts and integrating over spin-wave fluctuations, we
obtain equations of constraints, \bqa && \partial_{\mu}
J_{\mu}^{\sigma} + \sigma
\partial_{\mu} j_{\mu} \delta_{\mu\tau} = 0 , \eqa where $\sigma =
\pm$ represent spin $\uparrow$ and $\downarrow$, respectively.
These equations are solved to give \bqa && J_{\mu}^{\sigma} +
\sigma j_{\mu} \delta_{\mu\tau} = (\partial \times
\boldsymbol{c}^{\sigma})_{\mu} , \eqa where $c_{\mu}^{\sigma}$ is
U(1) gauge field to incorporate spin-wave excitations.

Integrating over $J_{\mu}^{\sigma}$ with Eq. (A3), we obtain \bqa
&& Z = \int D \phi_{\sigma}^{v}(\boldsymbol{r},\tau) D
a_{\mu}(\boldsymbol{r},\tau) D j_{\tau}(\boldsymbol{r},\tau) D
c_{\mu}^{\sigma}(\boldsymbol{r},\tau) \exp \Bigl[ -
\int_{0}^{\beta_r} d \tau \int d^{2} \boldsymbol{r} \Bigl\{
\frac{g_r}{2} [(\partial \times \boldsymbol{c}^{\sigma})_{\mu} -
\sigma j_{\mu} \delta_{\mu\tau}]^{2} - i (\partial \times
\boldsymbol{c}^{\sigma})_{\mu} (\partial_{\mu} \phi_{\sigma}^{v} -
a_{\mu}) \nn && + \frac{1}{2e_{a}^{2}} (\partial \times
\boldsymbol{a})^{2} \Bigr\} + \int_{0}^{\beta_r} d \tau \int d^{2}
\boldsymbol{r} \int_{0}^{\beta_r} d \tau' \int d^{2}
\boldsymbol{r}' \frac{1}{4c_{r} N_{F} \lambda^{2}}
j_{\tau}(\boldsymbol{r},\tau)
[\Pi(\boldsymbol{r}-\boldsymbol{r}',\tau-\tau')]^{-1}
j_{\tau}(\boldsymbol{r}',\tau') \Bigr] . \eqa It is
straightforward to perform the integration for $j_{\tau}$. Then,
we obtain a dual action in the first-quantization expression \bqa
&& Z = \int D \phi_{\sigma}^{v}(\boldsymbol{r},\tau) D
a_{\mu}(\boldsymbol{r},\tau) D
c_{\mu}^{\sigma}(\boldsymbol{r},\tau) \exp \Bigl[ -
\int_{0}^{\beta_r} d \tau \int d^{2} \boldsymbol{r} \Bigl\{
\frac{g_r}{2} (\partial \times \boldsymbol{c}^{\sigma})^{2} - i
(\partial \times \boldsymbol{c}^{\sigma})_{\mu} (\partial_{\mu}
\phi_{\sigma}^{v} - a_{\mu})+ \frac{1}{2e_{a}^{2}} (\partial
\times \boldsymbol{a})^{2} \Bigr\} \nn && - \int_{0}^{\beta_r} d
\tau \int d^{2} \boldsymbol{r} \int_{0}^{\beta_r} d \tau' \int
d^{2} \boldsymbol{r}' \frac{g_{r}^{2}}{2} [(\partial \times
\boldsymbol{c}^{\uparrow})_{\tau} - (\partial \times
\boldsymbol{c}^{\downarrow})_{\tau}](\boldsymbol{r},\tau) \nn &&
\Bigl\{ g_{r} \delta(\boldsymbol{r}-\boldsymbol{r}')
\delta(\tau-\tau') + \frac{1}{2c_{r} N_{F} \lambda^{2}}
[\Pi(\boldsymbol{r}-\boldsymbol{r}',\tau-\tau')]^{-1} \Bigr\}^{-1}
[(\partial \times \boldsymbol{c}^{\uparrow})_{\tau} - (\partial
\times \boldsymbol{c}^{\downarrow})_{\tau}](\boldsymbol{r}',\tau')
\Bigr] . \eqa We note that dynamics of spin-wave excitations is
strongly modified by nonlocal interactions from Fermi-surface
fluctuations of itinerant electrons through the Kondo-coupling
effect.

Considering the minimal coupling term of $c_{\mu}^{\sigma}
[J^{v}]_{\mu}^{\sigma}$ in Eq. (A5), where $[J^{v}]_{\mu}^{\sigma}
= (\partial \times \partial \phi_{\sigma}^{v})_{\mu}$ is
identified with a vortex current, one can construct the
second-quantization form of Eq. (A5), introducing vortex field
variables. As a result, we obtain an effective dual field theory
\bqa && Z = \int D \Phi_{\sigma}(\boldsymbol{r},\tau) D
a_{\mu}(\boldsymbol{r},\tau) D
c_{\mu}^{\sigma}(\boldsymbol{r},\tau) \exp \Bigl[ -
\int_{0}^{\beta_r} d \tau \int d^{2} \boldsymbol{r} \Bigl(
|(\partial_{\mu} - i c_{\mu}^{\sigma}) \Phi_{\sigma}|^{2} +
m_{v}^{2} |\Phi_{\sigma}|^{2} + \frac{u_{v}}{2}
|\Phi_{\sigma}|^{4} + \frac{g_r}{2} (\partial \times
\boldsymbol{c}^{\sigma})^{2} \nn && + i (\partial \times
\boldsymbol{c}^{\sigma})_{\mu} a_{\mu} + \frac{1}{2e_{a}^{2}}
(\partial \times \boldsymbol{a})^{2} \Bigr) - \int_{0}^{\beta_r} d
\tau \int d^{2} \boldsymbol{r} \int_{0}^{\beta_r} d \tau' \int
d^{2} \boldsymbol{r}' c_{r} N_{F} g_{r}^{2} \lambda^{2}
[(\partial \times \boldsymbol{c}^{\uparrow})_{\tau} - (\partial
\times \boldsymbol{c}^{\downarrow})_{\tau}](\boldsymbol{r},\tau)
\nn && \Pi(\boldsymbol{r}-\boldsymbol{r}',\tau-\tau') [(\partial
\times \boldsymbol{c}^{\uparrow})_{\tau} - (\partial \times
\boldsymbol{c}^{\downarrow})_{\tau}](\boldsymbol{r}',\tau') \Bigr]
. \eqa $\Phi_{\sigma}$ represents a meron field, which acts as a
source of $c_{\mu}^{\sigma}$ (spin-wave field), where the meron
current is given by $[J^{v}]_{\mu}^{\sigma} = - i
[\Phi_{\sigma}^{\dagger} (\partial_{\mu} \Phi_{\sigma}) -
(\partial_{\mu} \Phi_{\sigma}^{\dagger}) \Phi_{\sigma}]$.
$m_{v}^{2}$ is a mass of a meron, which identifies its quantum
critical point with $m_{v}^{2} = 0$. $u_{v}$ is a coupling
constant for its local self-interaction, determined from
non-universal short-distance physics.

As discussed in section II-B, we will not allow meron excitations.
Emergence of meron excitations has been discussed intensively in
Ref. \cite{DQCP}, where the interplay between Berry phase and
local self-interactions of merons would make "magnetic" monopole
excitations of compact U(1) gauge fields $a_{\mu}$ become
irrelevant in the renormalization group sense, giving rise to
merons as elementary excitations. Emergence of meron excitations
implies that of spinons, which carry fractional spin quantum
number $1/2$. On the other hand, we take the limit of $e_{a}
\longrightarrow \infty$, where such fractionalized excitations
cannot exist. Merons should be combined with anti-merons, forming
skyrmions, which means that spinons should be confined with
anti-spinons, resulting in spin $1$ excitations, identified with
conventional spin fluctuations. It is still extremely difficult to
clarify the precise condition for the emergence of deconfinement
although its existence seems to be accepted.

In order to realize the confinement ansatz, i.e., $e_{a}
\longrightarrow \infty$, we take the amplitude-frozen limit of
$\Phi_{\sigma} \sim e^{i\theta_{\sigma}}$ and obtain \bqa &&
(\partial_{\mu}\theta_{\sigma} - c_{\mu}^{\sigma})^{2} =
\frac{1}{2} \Bigl([\partial_{\mu}\theta_{\uparrow} +
\partial_{\mu}\theta_{\downarrow}] - [c_{\mu}^{\uparrow} +
c_{\mu}^{\downarrow}]\Bigr)^{2} + \frac{1}{2}
\Bigl([\partial_{\mu}\theta_{\uparrow} -
\partial_{\mu}\theta_{\downarrow}] - [c_{\mu}^{\uparrow} -
c_{\mu}^{\downarrow}] \Bigr)^{2} . \eqa At the same time, we
rewrite the kinetic energy of the vortex gauge field (spin-wave)
as follows, \bqa && \frac{g_r}{2} (\partial \times
\boldsymbol{c}^{\sigma})^{2} = \frac{g_{r}}{4} \Bigl([(\partial
\times \boldsymbol{c}^{\uparrow}) + (\partial \times
\boldsymbol{c}^{\downarrow})] \Bigr)^{2} + \frac{g_{r}}{4}
\Bigl([(\partial \times \boldsymbol{c}^{\uparrow}) - (\partial
\times \boldsymbol{c}^{\downarrow})] \Bigr)^{2} . \eqa Then, one
can rewrite Eq. (A6) in terms of newly defined vortices and gauge
fields, \bqa && {\cal L}_{v} = |(\partial_{\mu} - i c_{\mu}^{+})
\Phi_{+}|^{2} + m_{+}^{2}|\Phi_{+}|^{2} + \frac{u_{+}}{2}
|\Phi_{+}|^{4} + i a_{\mu} \epsilon_{\mu\nu\lambda}
\partial_{\nu} c_{\lambda}^{+} + \frac{g_{r}}{4}
(\partial \times \boldsymbol{c}^{+})^{2} + \frac{1}{2e_{a}^{2}}
(\partial \times \boldsymbol{a})^{2} \nn && + |(\partial_{\mu} - i
c_{\mu}^{-}) \Phi_{-}|^{2} + m_{-}^{2}|\Phi_{-}|^{2} +
\frac{u_{-}}{2} |\Phi_{-}|^{4} + \frac{g_{r}}{4} (\partial \times
\boldsymbol{c}^{-})^{2} \nn && + c_{r} N_{F} g_{r}^{2} \lambda^{2}
\int_{0}^{\beta_r} d \tau' \int d^{2} \boldsymbol{r}' [\partial
\times \boldsymbol{c}^{-}(\boldsymbol{r},\tau)]_{\tau}
\Pi(\boldsymbol{r}-\boldsymbol{r}',\tau-\tau') [\partial \times
\boldsymbol{c}^{-}(\boldsymbol{r}',\tau')]_{\tau} , \eqa where
$\theta_{\pm} = \frac{\theta_{\uparrow} \pm
\theta_{\downarrow}}{\sqrt{2}} \longrightarrow \Phi_{\pm}$ is a
''new" vortex field and $c_{\mu}^{\pm} = \frac{c_{\mu}^{\uparrow}
\pm c_{\mu}^{\downarrow}}{\sqrt{2}}$ is also a ''new" vortex-gauge
field. An essential aspect of Eq. (A9) is that only the $\Phi_{+}$
sector contains interaction with compact U(1) gauge fluctuations
$a_{\mu}$ while the dynamics of $\Phi_{-}$ decouples from the
gauge dynamics completely. If we take $e_{a} \longrightarrow
\infty$, $\Phi_{+}$ cannot appear as elementary excitations. This
is the confinement ansatz, which do not allow such fields as carry
the U(1) internal gauge charge of $a_{\mu}$. This dual language
can be interpreted as the fact that either spinons or anti-spinons
cannot appear as the physical spectrum.

Integrating over $a_{\mu}$ in the limit of $e_{a} \longrightarrow
\infty$, we see that $\Phi_{+}$ and $c_{\mu}^{+}$ disappear in the
physical spectrum. As a result, we reach an effective field theory
for skyrmion dynamics in the presence of itinerant electrons \bqa
&& {\cal L}_{sk} = |(\partial_{\mu} - i c_{\mu}^{-}) \Phi_{-}|^{2}
+ m_{-}^{2}|\Phi_{-}|^{2} + \frac{u_{-}}{2} |\Phi_{-}|^{4} +
\frac{g_{r}}{4} (\partial \times \boldsymbol{c}^{-})^{2} \nn && +
c_{r} N_{F} g_{r}^{2} \lambda^{2} \int_{0}^{\beta_r} d \tau' \int
d^{2} \boldsymbol{r}' [\partial \times
\boldsymbol{c}^{-}(\boldsymbol{r},\tau)]_{\tau}
\Pi(\boldsymbol{r}-\boldsymbol{r}',\tau-\tau') [\partial \times
\boldsymbol{c}^{-}(\boldsymbol{r}',\tau')]_{\tau} . \eqa

\section{Irrelevance of the Chern-Simons term at quantum criticality}

Starting from Eq. (4) and replacing $\Phi_{s}$ with
$\frac{1}{\sqrt{\mu_{sk}}} \Phi_{s}$, we obtain \bqa && Z_{sk} =
\int D \Phi_{s} D c_{\mu} \exp\Bigl[ - \int_{0}^{\beta_r} d \tau
\int d^{2} \boldsymbol{r} \Bigl\{ \Phi_{s}^{\dagger}
(\partial_{\tau} - i c_{\tau}) \Phi_{s} + \frac{1}{\mu_{sk}}
|(\partial_{\mu} - i c_{\mu} ) \Phi_{s}|^{2} +
\frac{m_{s}^{2}}{\mu_{sk}}|\Phi_{s}|^{2} +
\frac{u_{s}}{2\mu_{sk}^{2}} |\Phi_{s}|^{4} + \frac{1}{2q_{r}^{2}}
(\partial \times \boldsymbol{c})^{2} \Bigr\} \nn && -
\int_{0}^{\beta_r} d \tau \int d^{2} \boldsymbol{r}
\int_{0}^{\beta_r} d \tau' \int d^{2} \boldsymbol{r}' \frac{1}{2
p_{r}^{2}} [\partial \times
\boldsymbol{c}(\boldsymbol{r},\tau)]_{\tau}
\Pi_{\boldsymbol{r}\boldsymbol{r}',\tau\tau'} [\partial \times
\boldsymbol{c}(\boldsymbol{r}',\tau')]_{\tau} \Bigr] . \eqa

In order to introduce the statistical transmutation for skyrmion
excitations, we perform the Chern-Simons gauge transformation,
which attaches a flux to a skyrmion. Then, we obtain \bqa &&
Z_{sk} = \int D \Psi_{s} D c_{\mu} D \alpha_{\mu} \exp\Bigl[ -
\int_{0}^{\beta_r} d \tau \int d^{2} \boldsymbol{r} \Bigl\{
\Psi_{s}^{\dagger} (\partial_{\tau} - i c_{\tau} - i
\alpha_{\tau}) \Psi_{s} + \frac{1}{\mu_{sk}} |(\partial_{\mu} - i
c_{\mu} - i \alpha_{\mu}) \Psi_{s}|^{2} +
\frac{m_{s}^{2}}{\mu_{sk}}|\Psi_{s}|^{2} +
\frac{u_{s}}{2\mu_{sk}^{2}} |\Psi_{s}|^{4} \nn && +
\frac{1}{2q_{r}^{2}} (\partial \times \boldsymbol{c})^{2} - i
\frac{\theta}{2\pi} \alpha_{\mu} \epsilon_{\mu\nu\lambda}
\partial_{\nu} \alpha_{\lambda} \Bigr\} - \int_{0}^{\beta_r} d \tau \int d^{2}
\boldsymbol{r} \int_{0}^{\beta_r} d \tau' \int d^{2}
\boldsymbol{r}' \frac{1}{2 p_{r}^{2}} [\partial \times
\boldsymbol{c}(\boldsymbol{r},\tau)]_{\tau}
\Pi_{\boldsymbol{r}\boldsymbol{r}',\tau\tau'} [\partial \times
\boldsymbol{c}(\boldsymbol{r}',\tau')]_{\tau} \Bigr] , \eqa where
$\Psi_{s}$ is a fermionized skyrmion field and $\alpha_{\mu}$ is
the Chern-Simons gauge field to guarantee the statistical
transmutation. $\theta$ is the Chern-Simons angle, where $\theta =
\pi$ describes the fermion-boson or boson-fermion transmutation.

Shifting the skyrmion gauge field (spin wave) as $c_{\mu}
\longrightarrow c_{\mu} - \alpha_{\mu}$, we obtain \bqa && Z_{sk}
= \int D \Psi_{s} D c_{\mu} D \alpha_{\mu} \exp\Bigl[ -
\int_{0}^{\beta_r} d \tau \int d^{2} \boldsymbol{r} \Bigl\{
\Psi_{s}^{\dagger} (\partial_{\tau} - i c_{\tau}) \Psi_{s} +
\frac{1}{\mu_{sk}} |(\partial_{\mu} - i c_{\mu}) \Psi_{s}|^{2} +
\frac{m_{s}^{2}}{\mu_{sk}}|\Psi_{s}|^{2} +
\frac{u_{s}}{2\mu_{sk}^{2}} |\Psi_{s}|^{4} \nn && +
\frac{1}{2q_{r}^{2}} (\partial \times \boldsymbol{c} - \partial
\times \boldsymbol{\alpha})^{2} - i \frac{\theta}{2\pi}
\alpha_{\mu} \epsilon_{\mu\nu\lambda}
\partial_{\nu} \alpha_{\lambda} \Bigr\} \nn && - \int_{0}^{\beta_r} d \tau \int d^{2}
\boldsymbol{r} \int_{0}^{\beta_r} d \tau' \int d^{2}
\boldsymbol{r}' \frac{1}{2 p_{r}^{2}} [\partial \times
\boldsymbol{c}(\boldsymbol{r},\tau) - \partial \times
\boldsymbol{\alpha}(\boldsymbol{r},\tau)]_{\tau}
\Pi_{\boldsymbol{r}\boldsymbol{r}',\tau\tau'} [\partial \times
\boldsymbol{c}(\boldsymbol{r}',\tau') - \partial \times
\boldsymbol{\alpha}(\boldsymbol{r}',\tau')]_{\tau} \Bigr] . \eqa
Integrating over the Chern-Simons gauge field, we reach the
following expression \bqa && Z_{sk} = \int D \Psi_{s} D c_{\mu}
\exp\Bigl[ - \int_{0}^{\beta_r} d \tau \int d^{2} \boldsymbol{r}
\Bigl\{ \Psi_{s}^{\dagger} (\partial_{\tau} - i c_{\tau}) \Psi_{s}
+ \frac{1}{\mu_{sk}} |(\partial_{\mu} - i c_{\mu}) \Psi_{s}|^{2} +
\frac{m_{s}^{2}}{\mu_{sk}}|\Psi_{s}|^{2} +
\frac{u_{s}}{2\mu_{sk}^{2}} |\Psi_{s}|^{4} + \frac{1}{2q_{r}^{2}}
(\partial \times \boldsymbol{c})^{2} \Bigr\} \nn && -
\int_{0}^{\beta_r} d \tau \int d^{2} \boldsymbol{r}
\int_{0}^{\beta_r} d \tau' \int d^{2} \boldsymbol{r}' \frac{1}{2
p_{r}^{2}} [\partial \times
\boldsymbol{c}(\boldsymbol{r},\tau)]_{\tau}
\Pi_{\boldsymbol{r}\boldsymbol{r}',\tau\tau'} [\partial \times
\boldsymbol{c}(\boldsymbol{r}',\tau')]_{\tau} -
\mathcal{S}_{irr.}[c_{\mu}(\boldsymbol{r},\tau)] \Bigr] , \eqa
where $\mathcal{S}_{irr.}[c_{\mu}(\boldsymbol{r},\tau)]$ describes
the dynamics of $c_{\mu}$, which results from the integration of
the Chern-Simons gauge field. It turns out to be \bqa &&
\mathcal{S}_{irr.}[c_{\mu}(\boldsymbol{r},\tau)] \propto
c_{\mu}(\boldsymbol{q},i\Omega)
\Pi_{\mu\nu}(\boldsymbol{q},i\Omega)
c_{\nu}(-\boldsymbol{q},-i\Omega) \propto |\boldsymbol{q}|^{3}
c_{\mu}(\boldsymbol{q},i\Omega)
P^{T}_{\mu\nu}(\boldsymbol{q},i\Omega)
c_{\nu}(-\boldsymbol{q},-i\Omega) , \eqa where
$P^{T}_{\mu\nu}(\boldsymbol{q},i\Omega)$ is the projection
operator to the transverse direction. As shown in this expression,
the proportionality to $|\boldsymbol{q}|^{3}$ implies the
irrelevance of the Chern-Simons interaction at quantum
criticality.
\end{widetext}

\end{document}